*Original Article*

# Medical Data Asset Management and an Approach for Disease Prediction using Blockchain and Machine Learning

K. Shruthi[1], A. S. Poornima[2]

[1,2]*Department of Computer Science and Engineering, Siddaganga Institute of Technology, Karnataka, India.*

[1]*Corresponding Author : shruthik@sit.ac.in*

*Abstract - In the present medical services, the board, clinical well-being records are as electronic clinical record (EHR/EMR) frameworks. These frameworks store patients' clinical histories in a computerized design. Notwithstanding, a patient's clinical information is gained in a productive and ideal way and is demonstrated to be troublesome through these records. Powerlessness constantly prevents the well-being of the board from getting data, less use of data obtained, unmanageable protection controls, and unfortunate information resource security. In this paper, we present an effective and safe clinical information resource, the executives' framework involving Blockchain, to determine these issues. Blockchain innovation facilitates the openness of all such records by keeping a block for each patient. This paper proposes an engineering utilizing an off-chain arrangement that will empower specialists and patients to get records in a protected manner. Blockchain makes clinical records permanent and scrambles them for information honesty. Clients can notice their wwell-being records, yet just patients own the confidential key and can impart it to those they want.*

*Smart contracts likewise help our information proprietors to deal with their information access in a permission way. The eventual outcome will be seen as a web and portable connection point to get to, identify, and guarantee high-security information handily. In this adventure, we will give deals with any consequences regarding the issues associated with clinical consideration data and the chiefs using AI and Blockchain. Removing only the imperative information from the data is possible with the use of AI. This is done using arranged estimations. At the point when this data is taken care of, the accompanying issue is information sharing and its constancy. This is where Blockchain comes into the picture. Understanding Blockchain development guarantees that data is real and trades are secure. Blockchain development could work on clinical benefits by setting patients at the point of convergence of the clinical consideration structure and extending the insurance and interoperability of prosperity data. This paper is based in a general sense on dealing with clinical benefits data the board issues using Blockchain development and including a couple of key AI components. The fundamental thought process is to bring the attributes of AI and Blockchain together. AI assumes a pivotal part in identifying lethal illnesses. Then again, Blockchain innovation can reform clinical information base interoperability and limit unapproved record admittance. This would guarantee that the touchy patient information is firmly gotten. Expects to construct a safe, ML-driven medical care executive's framework that would guarantee that the sicknesses are precisely anticipated and sorted in the beginning phase.*

*Further, it guarantees that the prepared model channels the information and disposes of the multitude of individual subtleties of the patient and safeguards it from information holes and breaks. It drives the framework with Blockchain to get the exchanges among patients and the approved specialist. It also gives patients the adaptability to pick which specialist should see their wwell-being record and who should not.*

*Keywords - Blockchain, Ethereum, Flask, Ganache, Machine learning.*

## 1. Introduction

One of the biggest problems in the global health and medical sector is the inability to obtain medical data efficiently and on time. All Electronic Medical records or Electronic Health records (EMR/EHR) systems assume that each patient sees a doctor in a single clinic or the general state or province. This system only focuses on one practitioner, and using that information is neither useful nor accurate. Patients visit several doctors each year for various reasons. A wider aspect of the practitioner that is not usually considered is the chiropractor, pharmacist, etc. Patients also travel on vacation, work, and even move for long periods. Many patients are very interested in recording their health status with portable medical devices. This patient-generated health data is easily collected by providers or devices but never by doctors. As a result, it has proven difficult for patients to access their many records through the combined EMR due to security and privacy concerns. Even though patients are the legal owners of their medical records, and despite heavy medical investments by practitioners, healthcare institutions, and governments, healthcare management has always been hampered by the inability to obtain information, poor use of obtained information, poor record security, and privacy controls, which cannot be arranged. This is the main reason for developing a health management system usBlockchain. Due to this security consideration, Blockchain is implemented in this health management system.

Blockchain is viewed as a changeless record that is decentralized, circulated, and stores information in blocks. Each block of the Blockchain is associated with its first block through the connection. Since every header of the block has the hash worth of the first one, it turns out to be computationally infeasible for an outsider to change any information in the Blockchain as the need might arise to be considered each hub of the Blockchain. This makes Blockchain a changeless and secure data set to store private data. As each hub is free in the Blockchain and no hub is better than others, it shapes a decentralized framework that is strong for taking care of such huge information like clinical records. To carry out hashing of pointers, we use the Merkle tree. Merkle trees are cryptographic hash pointers addressed as paired trees.

The development of a Merkle tree is finished by taking a hash of a couple of pieces of information as the leaf hub, and its result is hashed till the root hub. Merkle trees make exchange checks more straightforward and quicker in the Blockchain. However, since permissionless Blockchains are vulnerable to being problematic, permissioned chains are being executed as they stay away from costly agreement components and work in trusted conditions. Today we have a colossal measure of information accessible in each area, and with the coming of innovation accessible, giving answers to some problems is conceivable. In the ongoing scene, we go over new creations and executions consistently. Broad information measures are gathered in medical care associations utilized to control AI models and simulated intelligence-driven applications. While this is an advantage, there are sure genuine electronic attack worries that have arisen in clinical thought associations during recent years. Issues, for example, information breaks or disclosure of touchy data of patients, are the main issues. This issue could be addressed by building effective frameworks utilizing AI upheld by Blockchain.

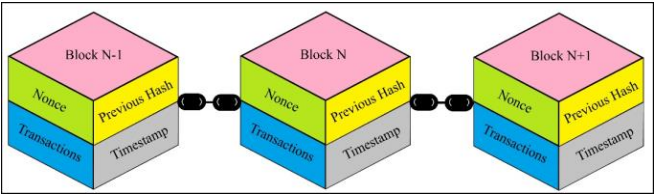

**Fig. 1 A standard structure of Blockchain**

### 1.1. Machine Learning in Healthcare

Using a predetermined set of ideas and technology, such as linguistic and statistical approaches, machine learning is a computer-based method for analyzing free-form text or voice to derive rules and patterns from the data. In terms of patient care, knowledge and experience are two very important factors for doctors to consider. However, people are restricted in their ability to learn new things by gathering data, whereas machine learning is quite good at it. A machine learning model that has been properly trained with a big collection of accurate training data lowers the need for human labor and effort by automating the jobs. The algorithm utilized and the volume and nature of data used for training are the major determinants of the model's effectiveness. It is preferable to use a training dataset that is as large as feasible. The training set's examples must be diverse to maintain harmony amongst the various target classes. The model will bias toward the target class with the majority of examples if the examples of the several target classes are unbalanced. The variety of training examples is crucial since it can be useful in situations when various testing data are seen. Supervised learning and unsupervised learning are the two categories of machine learning. Using trained algorithms, it is possible to just extract the pertinent data from the data. Utilizing trained algorithms, this is done. Large clinical datasets that doctors gather can be used to uncover hidden insights through data mining. Regularly this provides the specifics of the patient's symptom diagnosis, assisting them in choosing the appropriate course of treatment. An important set of disease-specific symptoms is considered, and a model is then constructed. Although relatively difficult, using machine learning to predict disease in healthcare, but very beneficial to clinicians.

### 1.2. Blockchain in Healthcare

Before Medical care, the patient's information was put away in a concentrated data set where it was exclusively open. The brought-together framework has many weaknesses, including centralization of force, essential issue disappointment, dishonest tasks, adaptability issues, trust issues, and protection issues. Patients' information ought to be open in a decentralized manner. The benefit of the conveyed data set is that every member has a duplicate of the data set. As a rule, controlling access and coordinating the frameworks and cycles in your duplicate of a data set is simpler. In any case, synchronization of changes to every data set is required. Taking care of disappointments and clashes can add intricacy and information respectability issues.

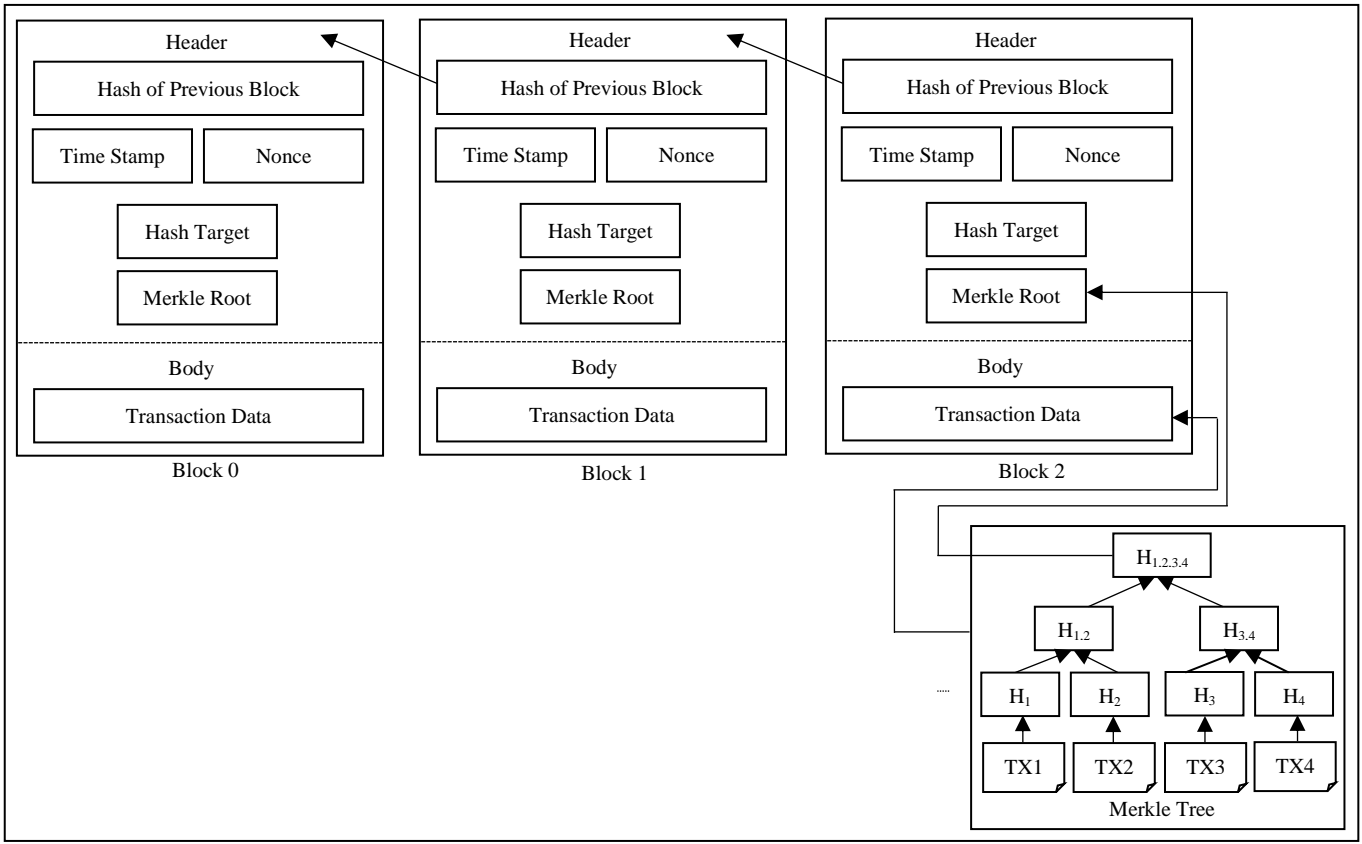

**Fig. 2 Standard block structure**

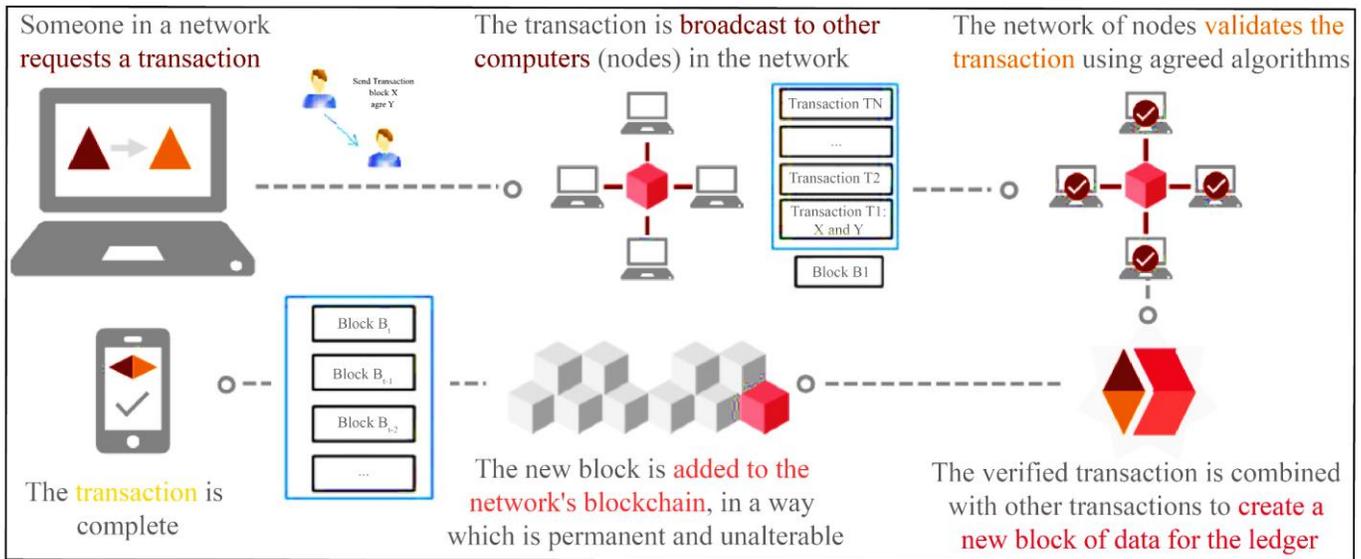

**Fig. 3 Diagram of the Transaction flow in Blockchain**

Blockchain is a decentralized, shared, add-just, circulated record where solitary exchanges are encoded into blocks by material encryption, added to the record, and never erased. The data in Blockchain is checked on a very basic level by a connected rundown of encoded trades that uses a hash. The hash capability produces a hash by scrambling the data taken care of in the Blockchain. It shapes the underpinning of a decentralized restorative help stage shared by the patients and providers, going about as a point of interaction with the patient's record. It comprises blocks, and these blocks are connected utilizing an unmodifiable key referring to the instrument.

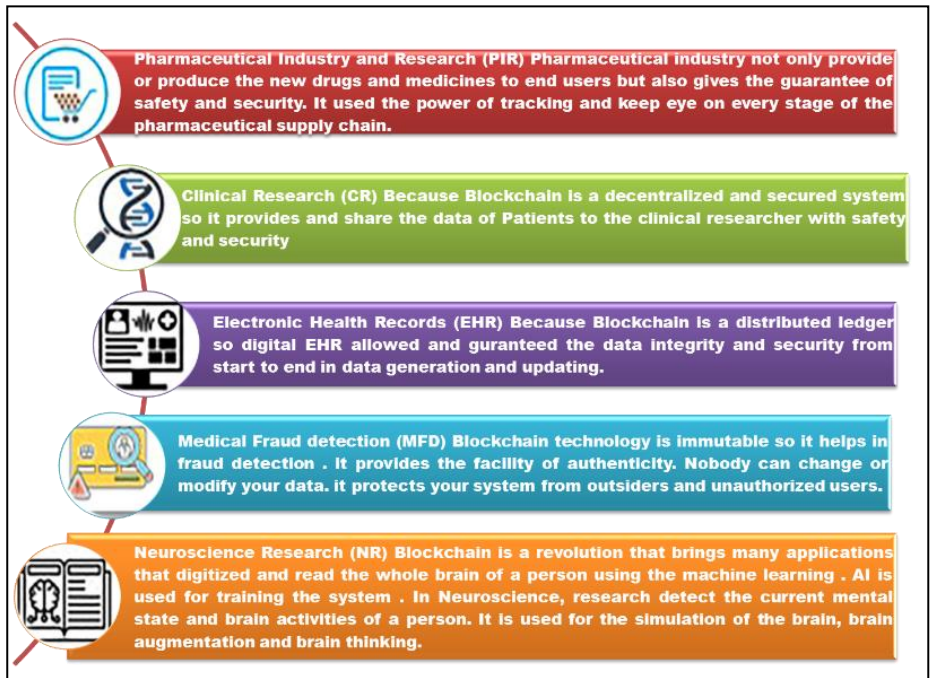

**Fig. 4 Applications of blockchain technology in healthcare**

Each quiet and specialist goes about as a hub; every single movement done by any hub will be recorded as an exchange. These exchanges are packaged together to shape a block, and diggers will mine these blocks, and blocks will be a piece of the Blockchain. In Blockchain, each block stores Keccak 256-cycle parent hash. The following block connects each block through the previous hash. These blocks store the data in a got way in an add-just data set.

### 1.3. Consensus in Blockchain

The agreement in Blockchain innovation ensures that information is real and exchanges are secure. Blockchain innovation might improve medical care the board by setting patients at the focal point of the medical care framework and expanding the protection and interoperability of well-being information. The essential spotlight lies in tackling medical care information the board issues by utilizing Blockchain innovation and including a few key elements utilizing AI. Since we are involving Ethereum Blockchain, the execution agreement component we are utilizing is Proof of Stake(PoS). Proof of Stake(PoS) is characterized as a digital currency agreement system for handling exchanges and making new blocks in a Blockchain as well as it is a technique for approving sections into a circulated decentralized Blockchain and keeping the Blockchain secure.

In Proof of Stake (PoS), excavators approve block exchanges given the number of coins a validator contributes as stakes. Proof of Stake(PoS) assists with approving a Blockchain and adding new blocks more safely than Proof of Work(PoW). Proof of stake (PoS) is viewed as safer concerning an assault on the organization. Proof of Stake(PoS) utilizes Merkle Patricia Tree, which is utilized to store information and works with productive supplement/erases activities and key-esteem queries are exceptionally proficient.

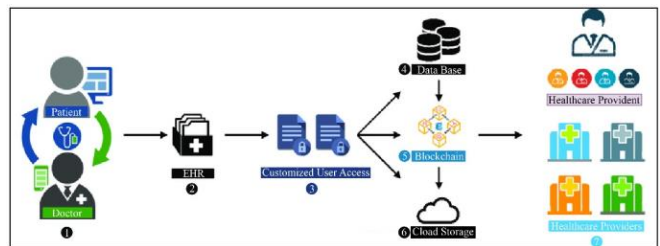

**Fig. 5 Blockchain-based healthcare data management system.**

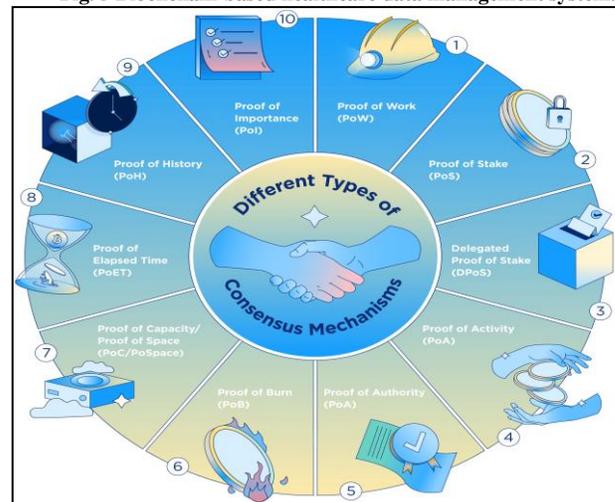

**Fig. 6 Consensus in Blockchain.**

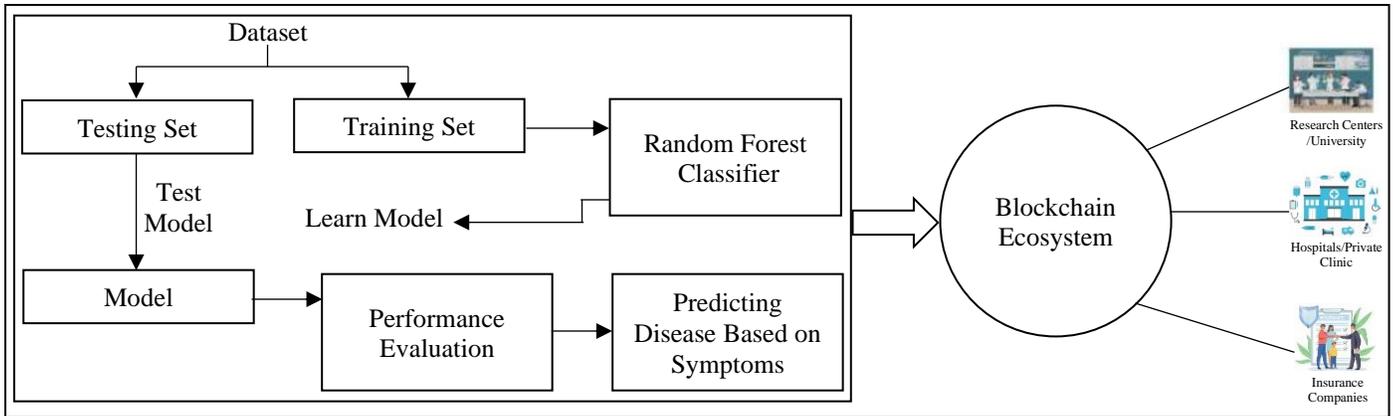

**Fig. 7 Blockchain Learning Architecture**

### 1.3.1. Proof of Stake Working

At the point when an exchange should be effectively finished, that total exchange should be shared by the validator, who will be liable for adding the exchange to the Blockchain. Validators, once persuaded blocks to be proposed, need to communicate to all close by excavators in one's organization, and later, they all utilize one calculation to propose new blocks. Around here, no less than 128 individuals ought to approve the exchange, the gathering is classified as "board", and the Byzantine issue is addressed. There will be a timestamp of 30sec to propose and approve the exchange inside that timestamp target. In Proof of Work, the digger who settles the mysterious riddles the quickest will mine the new block and get compensated for it. At the same time, different contenders are left with computational misfortune and nothing close by. Verification of stake disposes of this fault and thus opens the opposition to just those excavators who can stand to place the cash in stake. This guarantees natural selection and does not let any other person lose. So the validators do not have to utilize critical measures of computational power since they are chosen indiscriminately and are not contending. They do not have to mine blocks. They simply have to make blocks when picked and approve proposed blocks when they are not. This is known as verifying. On the off chance that a noxious block is confirmed, the validator loses his stake. It rules over Proof of Work as the last option has gotten everyone's eyes and has gained notoriety for the enormous measures of computational power and power it consumes.

### 1.4. Integrating Machine Learning and Blockchain

Blockchain and AI advances can moderate medical services issues like sluggish admittance to clinical information, unfortunate framework interoperability, absence of patient organization, and information quality and amount for clinical examination. Blockchain innovation works with and gets data capacity so that specialists can see a patient's whole clinical history. Yet, scientists see just factual information rather than any private data. AI can utilize this information to see themes and give exact expectations, offering more help for the patients and in research-related fields where there is a requirement for precise information to anticipate valid outcomes. The clinical history and patient's delicate information are put away in Blockchain. AI specialists and scientists can utilize the patient's clinical information to foresee the sickness's seriousness and treat it at the beginning phase. The AI models worked for a specific illness, taking the side effects and different boundaries as information and results from the seriousness of the sickness regarding characterization.

### 1.4.1 Uses of dApps and Kinds of Blockchains

One of the most important new areas of software development is "decentralized applications," or "dApps," which are programs that use blockchain technology. In July 2021, Google Scholar searched for scientific and technical documents and returned 36,700 results for developing "smart contracts." This number was either significantly higher or equal to the results for the development of microservices (20,500), global software engineering (7670), development of DevOps (23,500), and even "software development" for the Internet of Things (30,400). In addition to the SCs running on a blockchain, dApps also include the software that manages data outside of the Blockchain and the user interface for interacting with that software. In the beginning, SCs were primarily used to manage second-level digital currencies, or "tokens," which were primarily used to finance Initial Coin Offerings (ICOs), which are crowdfunding operations that gather cryptocurrencies to finance startups [25]. Data notarization, finance and insurance contracts, supply chain management, smart and microgrid management, and the health sector (personal records, pharmaceutical product delivery, clinical trials, etc.) are just a few of the many applications for which dApps are currently being used in addition to tokens [26–28], [30], identity management and access control systems [28], a decentralized notary [24], gambling, gaming, and voting [28], and numerous other alternatives [29,31]. Automated contractual obligations can be enforced using dApps and SCs without having to rely on a central authority and without having to worry about space or time. In a nutshell, the following characteristics distinguish a blockchain-based system:

- Dissemination: The system is resilient and secure because the data is stored on multiple computers.
- Continuity: Every aspect of a transaction can be retraced, and the original address of each one can be precisely identified.
- Trust: Asymmetric cryptography ensures ownership of an address where assets can be stored.
- Isolation: To send a transaction, the addresses' owners do not need to publish their names; they only need to demonstrate that they own the associated private key.
- Decentralization: transactions are managed without a single point controller.
- Transparency: Blockchain's contents can be verified and easily accessed.
- Immutability: There is no longer any way to alter the accepted data in any way.
- Programmability: Programmable, fully verifiable code and execution for complex actions (smart contracts) are also available.
- Low-cost: Open-source software manages the system, which has low operating and maintenance costs (depending on transaction fees).

Structure of the paper. The remainder of this paper is organized as follows. Section 2 presents the related work; Section 3 discusses the framework to choose the blockchain platform best suited for our purposes, describes and proposes dApp architecture; section 4 draws the paper's results; and discussion Section 5 draws the paper's conclusions.

## 2. Literature Survey

One research paper by Kai Fan, Shangyang Wang, Yanhui Ren, Hui Li, and Yintang Yang titled "MedBlock: Efficient and Secure Medical Data Sharing Via Blockchain" [1] aims to provide a familiar method of storing patient information at hospitals. For a common patient, data is stored in multiple hospitals. Thus for a patient to get their multiple records accessible through a summarized EMR is proven hard because of security and privacy concerns. This paper provides a solution to the above problems by giving a blockchain-based information management system called 'MedBlock'. Here MedBlock's distributed ledger provides easy retrieval and access to data in electronic medical records. This paper also achieves an improved consensus mechanism and high data security by utilizing symmetric cryptography and customized access control protocols.

Table 1. Classification of blockchain types concerning validation and access

| Action | Blockchain Type | | |
|---|---|---|---|
| | **Public** | **Permissioned, Open** | **Permissioned, closed** |
| **Managing the right to add a node** | Not considered | Based on the original legal agreement | Based on the original legal agreement |
| **Adding a node able to mine/validate** | Everyone will pay a high price for mining. | Only if given permission, possibly through smart contract voting, All | Only with permission |
| **Adding a node holding Blockchain** | Everyone | Everyone | Only with permission |
| **Deploying a smart contract** | Everybody, for a fee or "gas." | Only with permission | Only with permission |
| **Sending transactions able to change the state (writing rights)** | Paying a fee or "gas" for everyone. The majority of smart contracts will only alter their state if transactions originate from authorized addresses. | The state of every smart contract will only change if transactions come from authorized addresses. | Only if you are properly authorized and logged in, additionally, the address could be checked. |
| **Sending read-only transactions to one or more smart contracts** | Everyone | Everyone, but only transactions from authorized addresses, will be accepted for the request. | Only if you are properly authorized and logged in, additionally, the address could be checked. |
| **Reading the content of Blockchain** | Everyone | Everyone | Only if properly logged in and granted permission |

**Table 2. The features needed by a dApp system and how public and permissioned blockchains support them.**

| Sl. No | Feature | Description | PubliBBlockchainnn | PermissioBlockchain |
|---|---|---|---|---|
| 1 | **Immutability** | A blockchain is an append-only system. Once written, the information cannot be altered or removed. The programs and data running on Blockchain must be verifiable, immutable, and counterfeit-proof | Very high | Very High |
| 2 | **Transparency** | The data and the work performed on Blockchain must be traceable. | Very high | It can vary greatly depending on the system. |
| 3 | **Trust** | Even in the absence of trustworthy participants, a blockchain can guarantee trust. | Even if participants lack mutual trust, it works well. | The initiative cannot succeed without trust, and Blockchain can withstand attacks from a small number of participants. |
| 4 | **Identity** | All writing on Blockchain must originate from specific sources. | Strongly based on ownership of private keys, owners can publicly associate their addresses with their identities. | Strong if username and password are used, and extremely strong if private key ownership is used. |
| 5 | **Historical records** | Blockchain and other system repositories and applications must continue to function for an appropriate amount of time, typically years or decades. | Based on the miners' reward, very high | High depending on the validators' willingness and ease of use. |
| 6 | **Ecosystem** | Is interoperability between partners supported by the architecture, as opposed to a single company system? | Achieved | Easily achieved |
| 7 | **Efficiency** | Even with a large number of users and transactions per unit of time, the system should be able to handle the required throughput and response time. | Very few transactions occur every second. | There may be a lot of transactions per second. |
| 8 | **Privacy** | Only known users should be granted access to Blockchain, including the ability to alter its state, and this permission should be granted at various access levels. | Very few smart contracts can allow actions based on the particular address that sends the transaction. | High can be enacted at several levels. |
| 9 | **Scalability** | If necessary, the system ought to be able to expand. | Poor scalability if there are more dApps and users. | By splitting the nodes or putting additional blockchains on the same node, or both, |
| 10 | **Cost** | To accommodate the size of the dapp and the number of transactions per second, the blockchain system should be open source, simple to deploy, and only require a limited amount of hardware and network bandwidth resources. The price should not fluctuate. | There are only costs associated with software execution, which can be extremely volatile. | Execution costs are typically low and predictable, as are infrastructure costs. |
| 11 | **Deployment Costs** | The cost of deploying the system is low. | Typically, adding a node is not expensive. | A node can be added for very little money. |
| 12 | **Development Costs** | The cost of deploying the system is low. | Costs are determined by the maturity of development tools and the availability of developers. | Costs are determined by the maturity of development tools and the availability of developers. |

Catering to the needs of issues that arise in remote patient monitoring systems, Kristen N. Griggs, Olya Ossipova, Christopher P. Kohlios, Alessandro N. Baccarini, Emily A. Howson and Thaier Hayajneh published a paper titled "Healthcare Blockchain System Using Smart Contracts for Secure Automated Remote Patient Monitoring" [2] for resolving security concerns related to exchange and recording of data transactions. To resolve these issues, this paper proposed a blockchain-based smart contract to ensure safe analysis and medical sensor management. The paper used the Ethereum protocol and created a system where the smart device communicates with the sensors and calls smart contracts while at the same time logging every event on Blockchain. This benefits patients with real-time tracking and intervention in the medical field by notifying patients and practitioners while simultaneously providing security. This paper resolves all security issues in smart medical device health management HIPAA-compliantly.

Highlighting flaws in the centralized approach to maintaining health records, the research paper "Health Record Management through Blockchain Technology" by V. M. Harshini, S. Danai, H. R. Usha, and M. R. Kounte [3] brings out the issue of a data breach. There is a huge loss for institutions and patients when a data breach happens, in terms of monetary and health loss. Moving forward, this paper also reveals the flaws in institution-centralized data control and gives a decentralized system usBlockchain as a solution. In health care, smart contracts help to make things simpler. Where Blockchain will be used for invocation, record production, and validation, the proposed solution could be used for various problems in the medical domain, such as record maintenance, record sharing, billing, medical research, etc.

It lays the theoretical base for a model of maintaining records with the helpBlockchain, further research, and practical implementation. Similarly, research titled "A Secure and Scalable Data Source for Emergency Medical Care using Blockchain Technology" is a paper that focuses on the poor management and causes of failure of Emergency Medical care, written by S. Hasavari and Y. T. Song [4] A lot of pre-hospital deaths are caused due to the lack of medical history of patients. A lot of patients' data is generated in some other hospital and later is relevant to some other hospital. It is very difficult for a different healthcare system to retrieve a patient's medical history from the previous healthcare system. In this paper, a safe filing system Blockchain is proposed as the remedy for emergency access to patients' records. All medical records concerns like authentication and privacy have been taken care of in this approach. With the help of the practical implementation of this paper, the ambulance crews can access patients' medical history and avail high-quality pre-hospital care, which leads to a much lower death rate.

Blockchain in the medical system has few healthcare systems providing targeted sharing protocols for medical and personal health data. The paper titled "MedBlock: Blockchain-based Multi-role Healthcare Data Sharing System" by Y. Yu, Q. Li, Q. Zhang, W. Hu, and S. Liu [5] proposes a data management system for multiple users by combinBlockchain and the InterPlanetary File System (IPFS). Another work in similar literature, "Blockchain-based approach for e-health data access management with privacy protection" by L. Hirtan, P. Krawiec, C. Dobre, and J. M. Batalla [6] provides a design whBlockchain is used in the healthcare system for the storage of information in clinics, and hospitals based on the patient's access policies.

The paper shows 2 types of chains, i.e. a private sidechain and a public main chain, which either keep information on the patients' real ID or information with a temporary ID. Research, "Blockchain in Healthcare: A Patient-Centered Model", by Hannah S Chen, Juliet T Jarrell, Kristy A Carpenter, and David S Cohen, Xudong Huang [7] proposes a technology that could be utilized in many places for sharing and storing health records for both hospitals and application interfaces. AfBlockchain gained popularity in software research. It started providing patient data authority. This project thus created a change in authority for patients resulting in a patient-centered model.

A similar work titled "MedAccess: A Scalable Architecture for Blockchain-based Health Record Management" by M. Misbhauddin, A. AlAbdulatheam, M. Aloufi, H. Al-Hajji, and A. AlGhuwainem, [8] aims for a decentralized medical structure usBlockchain. However, the price of storing information and records is overpriced on Blockchain. This price is due to the high amount of computations required in a transaction before committing to Blockchain. The cost of managing patients' electronic health records (EHRs) is not feasible and costly on a blockchain service. Here, a project design solution is provided to reduce the high expenses on a blockchain. Furthermore, the usage of a watermark as a way to store patients' media results in addition to the usage of encryption on IPFS objects stored in the network.

## 3. Methodology

### 3.1. Blockchain

A Blockchain is a developing rundown of records, called connected blocks utilizing cryptography. Each block contains a cryptographic hash of the past block, a timestamp, and exchange information (by and large, addressed as a Merkle tree). The timestamp confirms that the exchange information was there at the time the block was delivered, permitting it to be hashed. The blocks structure a chain, with each new block supporting the first ones as it contains the data of the past block.

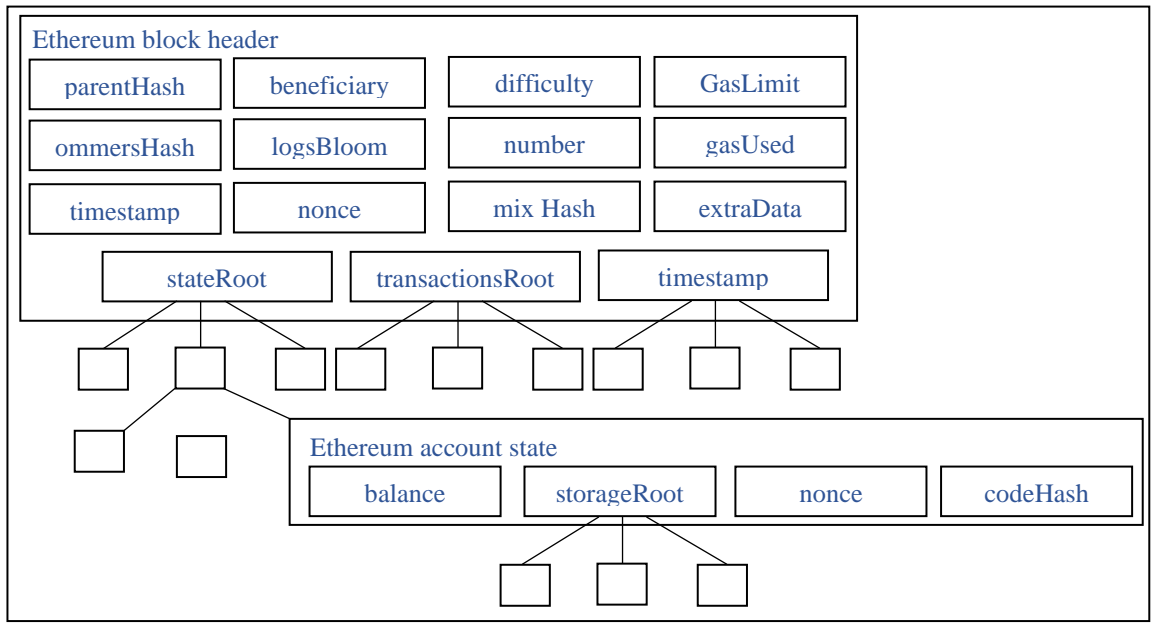

**Fig. 8 Ethereum Block Structure**

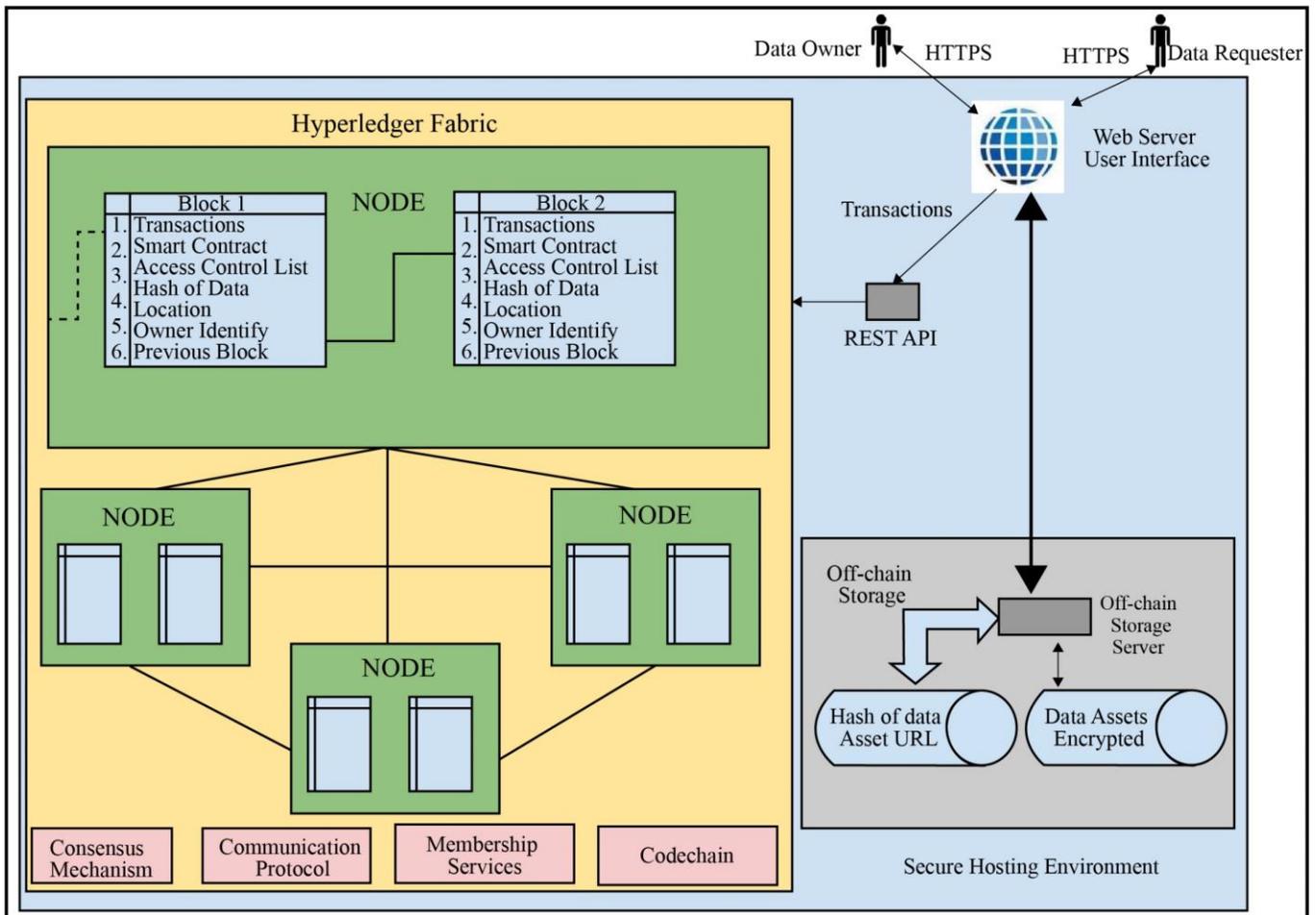

**Fig. 9 Medical Data Asset Management Architecture**

Thus, Blockchain is impervious to information altering since the information in some random block, once recorded, cannot be changed reflectively without influencing every ensuing block. Hyperledger Texture, an open-source project from the Linux establishment, is the measured Blockchain system and true norm for big business Blockchain stages. Blockchain Innovation: A Blockchain is a decentralized, circulated, and periodically open, computerized record comprising of records called blocks that are utilized to record exchanges across numerous PCs so that any elaborate block cannot be changed retroactively without the modification of every single ensuing block. Ethereum: Ethereum is a stage fueled by Blockchain innovation that is most popular for its local cryptographic money, called ether, or ETH, or essentially Ethereum. The conveyed idea of Blockchain innovation makes the Ethereum stage secure, and that security empowers ETH to accumulate esteem.

Block is the center essential unit of Blockchain. Blocks contain exchanges and some significant data, for example, past hash, which guarantees unchanging nature and security in the Blockchain organization. Each block contains a past hash successively, so altering the data is practically infeasible. The block will be made when a digger finds nonce, which is a one-time produced esteem after a ton of calculation. It is important to send secret phrases safely and forestall replay attacks.

As displayed in Figure 9, we have two principal elements, the information proprietor and the information requester, who might be taking care of the information. The reason for the thought is to give the proprietor complete command over their information. Subsequently, we will utilize Discretionary Access Control (DAC) strategy. The design depends on Hyperledger texture, which contains various hubs to store clinical history. Hyperledger Texture is a disseminated record stage that is permissioned where all hubs in the organization have a personality. Every hub has a hash key which is utilized to get information from off-chain capacity safely. Every patient (the information proprietor) has its Blockchain. Admittance to the Blockchain can be given to the information requester, similar to specialists, drug stores, and medical clinics, for a predetermined timeframe. The patient can give an extremely durable agree too to a requester to work with crisis admittance to recover information. To keep up with the trustworthiness of the information, the patient will have the choice to check assent subtleties given till a specific moment. The application would likewise work with the bringing of reports and solutions both by the patient as well as the information requestors. The information requestors who have assent from the patient will have the authorization to add the blocks to the chain. The specialist can allude to tests and medicines that the patient needs to undergo and endorse drugs utilizing the Blockchain.

Moreover, the specialist can add remarks and media to the blocks. At the point when a patient visits a clinic, another solution is made. Likewise, emergency clinics will have the consent to make another report on understanding Blockchain. Furthermore, clinics will have the usefulness to enlist a patient for the treatment suggested by the specialist and begin the treatment.

### 3.2. API Workflow

The flow chart shown in Figure 10 gives the API workflow of the architecture. The description of each entity within the system is explained in Table 3.

### 3.3. Off-Chain Storage

The clinical information in the executive's framework will contain an enormous measure of information that could not be obliged on the actual block. To scale the framework, we would require capacity off the Blockchain. This capacity is called off-chain capacity. The off-chain capacity can be an organized or unstructured capacity containing various information from the text to pictures. There are different issues with current on-chain capacity: access issues, execution issues, security issues, achievement issues, and cryptographic issues. We can utilize an off-chain, shared and secure capacity to stay away from this large number. The undertaking will utilize a NoSQL data set named CouchDB to outfit the upsides of conveyed frameworks. CouchDB is a record-based data set that works with both organized as well as unstructured information. It additionally is profoundly viable with hyper ledger and exceptionally receptive to HTTP demands which makes the exchange more straightforward and quicker. The CouchDB likewise has an element of replication and follows the Corrosive property, which forestalls any sort of information misfortune and keeps up with the atomicity of every single exchange occurring in the chain. The blocks in Blockchain will contain a hash esteem which will refer to the partition where the information is put away in the data set.

### 3.4. Machine Learning
#### 3.4.1. Machine Learning Libraries

- Pandas - Python library for data manipulation and analysis.
- Numpy - Library that supports large dimensional array and matrix operations.
- Pickle - Module used for serializing and deserializing Python object structure.
- Seaborn - Data visualization library used for statistical graphs.
  Scikit-Learn - It features various classification, clustering, and regression algorithms.

The datasets for different sicknesses were taken from Kaggle. At first, information preprocessing is finished to supplant missing qualities, handle exceptions, and track down the connection between the properties. Then, at that point, unique AI calculations are applied to each dataset to conclude the best appropriate calculation for every illness expectation because of the exactness and disarray grid acquired for each. Python records containing code of best-anticipating calculation alongside the dataset are utilized for the further step of the task.

### 3.4.2. Flask

- Flask is a micro web framework written in Python.
- It is used to develop web applications easily.

Respond supports the front end of the application. The side effects of a patient are taken through the UI and are sent as POST solicitation to the backend. In the flagon application, the separate courses are characterized where the information from the patient is acknowledged and further worked to foresee the sickness. Each prepared model related to the expectation of a specific illness is changed over completely to a pkl document utilizing the "pickle" module. The test information sent from the front end is applied to these prepared models, and results are shown.

### 3.4.3. Algorithms Used
*Support Vector Machine*

It is a managed AI calculation utilized for both grouping and relapse. This calculation's fundamental benefit is finding the most ideal choice limit that can accurately isolate the greatest focuses into various classes. Among different conceivable choice limits, the best limit is called Hyperplane.

*Pseudocode*

```
Consider data point :X
Target label:Y
    if Y₁(W.X₁ + b) -1 =0:
        then (Xi, Yi) is support
vectors
        then save parameters W,b
    else if Y₁(W.X₁+b)-1>0:
        then save parameters W,b
    else if Y₁(W.X₁+b)-1<0:
then update parameters W,b
```

*K Nearest Neighbor*

It is a straightforward characterization calculation. Another information point is ordered based on closeness with the closest currently arranged places. Here, K is the quantity of closest neighbors considered. The greater part of casting a ballot is typically finished on the K focuses. It is a significant element in deciding the most reliable incentive for K. It considers the larger part of deciding in favor of characterization examples. It ascertains normal if there should aise an occurrence of regression.

As displayed in Figure 14, the fundamental data is gathered. Brilliant arrangements are written in a document with .sol expansion. It comprises the multitude of fields referenced for both Specialists and Patients in doctors. sol and patient. sol documents separately. Savvy contracts are ordered and sent. The conveyed agreements are shown in the ganache by dispensing a block for each agreement alongside exchange hash, contract address, block number, account, gas utilized, gas cost, esteem sent, and all out ethers deducted from the ganache.

### 3.4.4. Ganache
*Ganache is a personal Blockchain for rapid Ethereum, and Corda distributed application development. One can use Ganache across the entire development cycle, enabling one to develop, deploy, and test dApps in a safe and deterministic environment. Ganache comes in two flavors: a GUI and CLI.*

### 3.4.5. Metamask

MetaMask is a free crypto wallet software that can be connected to virtually any Ethereum-based platform. This will allow storing any assets created or bought and connecting to any platform built on top of the Ethereum Blockchain.

### 3.4.6. Moralis

Moralis is a leading web development platform offering everything the user needs to create, host, and grow great dApps in one place. It allows one to simply interface with an infinite number of external projects, chains, and technologies. Moralis is the quickest method for building and sending dApps on Ethereum, BSC, Polygon, Solana, and Elrond. All Moralis dApps are cross-chain, of course. Expanding on Moralis guarantees that dApp is future verification. Regardless of whether new Blockchains are developed, dApp will quickly chip away at any chain. Moralis gives a serverless data set where the crude information is put away, and the information is questioned in a representable configuration utilizing ordering. The shrewd agreements are composed utilizing the strength programming language for patient specialists to run on every hub of the Ethereum Blockchain. Each shrewd agreement will be arranged and conveyed in the neighborhood Blockchain. The nearby Blockchain utilized is Ganache which comprises 10 unique hubs with 100 Ethers each. The application is connected to the nearby Blockchain through the web3 javascript connector and the chrome expansion metamask. At whatever point a node(patient or specialist) signs in, the exchange will be recorded, and the action done by that hub will be put away as an exchange inside a block by every single hub of the Blockchain.

**Table 3. Terminology table**

| Notation | Description |
|---|---|
| Patient | The patient registers to the network requiring their Aadhar number (a 12-digit individual identification number of India) and a consenter ID. This provides temporary consent to the doctors based on their ID. |
| Hospital | The hospital user creates a report for patients about their tests, details, etc., using the patient ID. |
| Doctor | The doctor user refers tests to patients and the type of tests and treatment to be conducted. |
| PathLab | This user tests based on the supervisor's ID and prescribes drugs to the patient. |
| Pharmacies | Pharmacies give drugs along with a recommendation message. They refer treatments to patients accordingly. |
| Hospital | This user begins treatment based on the supervisor ID. The doctor further adds comments and media files to the patient's records. |

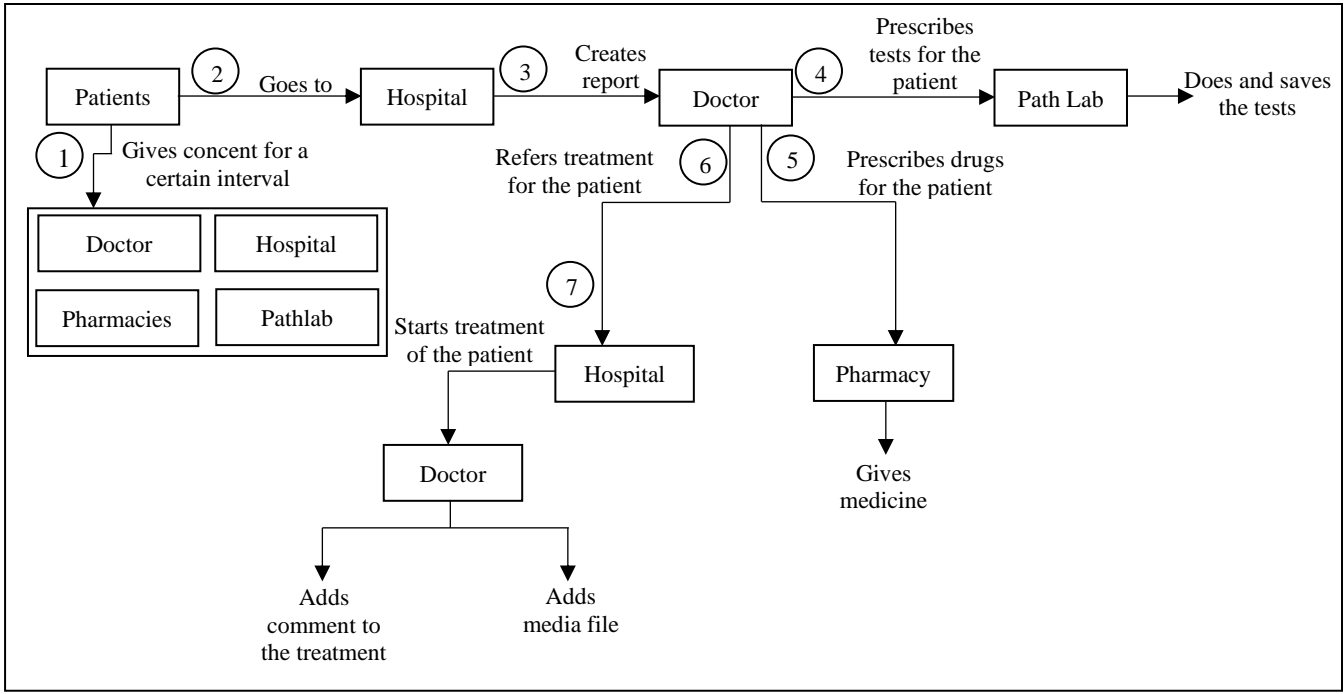

**Fig. 10 Data Flow within the system**

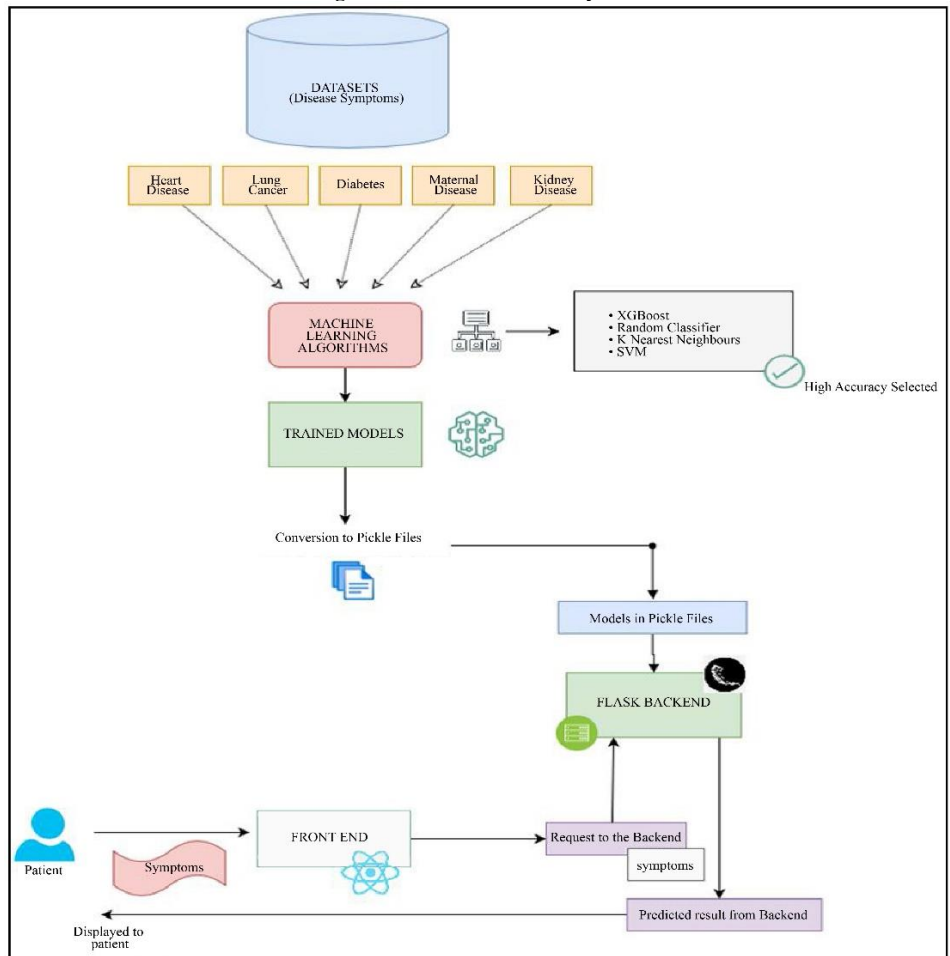

**Fig. 11 Disease Classification workflow using Machine Learning model**

**Pseudocode**

```
Consider a data point "x" to be classified.
1. Calculate d(x, xi) for i=1,2,...n where d indicates the Euclidean distance
between the points considered.
2. Represent the n distances in ascending order.
3. Consider first k distances from the list where k is a positive integer.
4. Determine the k-points corresponding to these k-distances.
5. Let kⱼ denote the number of points in the jᵗʰ class among the classes the
k points belong to, i.e. k >= 0.
6. If kᵢ > kⱼ and i != j, then put the query data point, x in class i.
Random Forest Classifier: It is an improved version of the decision tree
classifier. It contains many decision trees on various samples of the
available datasets and calculates the average to improve the prediction. It
can be used for classification as well as regression. It considers majority
voting for classification instances and calculates the average in case of
regression.
1. Store the prediction of each decision trees
P+=1 if X(i)=P
N+=1 if X(i)=N
2. Compute the total votes for an individual class.
3. Declare majority class as the final outcome
XGBoost Classifier: Extreme Gradient Boosting classifier helps in fast and
accurate classification.
Boosting is a technique to enhance the accuracy of a model using multiple
weak models. Sequential decision trees are created. Weights are assigned to
each input variable; these variables are then fed to the tree.
1. Load feature vector
2. Save the individual scores in S
3. Instantiate classifier as clf
4. fit clf
5. Generate FIs
6. Determine FI threshold FIp
for i=1 to n do
if(FI(Xi)>=FIp) then
append FI(Xi) into S
7. Use the scores in S to generate the feature vector
```

**Table 4. Machine learning dataset information**

| Disease | Attributes | Target class |
|---------|-----------|--------------|
| **Heart** | age, anaemia, creatinine_phosphokinase, diabetes, ejection_fraction, high_blood_pressure, platelets, serum_creatinine, serum_sodium, sex, smoking, time | Risk of heart: Low/High |
| **Lung Cancer** | gender, age, smoking, yellow_fingers, anxiety, peer_pressure, chronic_disease, fatigue, allergy, wheezing, alcohol_consuming, coughing, shortness_of_breath, swallowing_difficulty, chest_pain | Risk of Lung cancer: Low/High |
| **Maternal Health** | Age, SystolicBP, DiastolicBP, BS, BodyTemp, Heart Rate | Risk: Low/Mid/High |
| **Kidney** | id, age, bp, sg, al,su, rbc, pc, pcc, ba, bgr, bu, sc, sod, pot, hemo, pcv, wc, rc, htn, dm, cad, appet, pe, ane | Chronic Kidney Disease/ Not |
| **Diabetes** | Pregnancies, Glucose, BloodPressure, SkinThickness, Insulin, BMI, Diabetes Pedigree Function, Age | Diabetic/Not |

**Table 5. Blockchain property**

| Property | Value/Description |
|---|---|
| Consensus | Handled by a set of nodes |
| Transaction Validation | Set of Authorised nodes |
| Transaction Reading | Any node or set of predefined node |
| Data Immutability | Could be tampered |
| Transaction throughput | high |
| Network Scalability | Low to Medium |
| Infrastructure | Decentralized |
| Features | ● Applicable to tightly controlled business<br>● Fee-free transaction<br>● Laws on services are easier to manage<br>● Effective defense from outside perturbations |
| Example | HyperLedger, Ethermint, Tendermint |

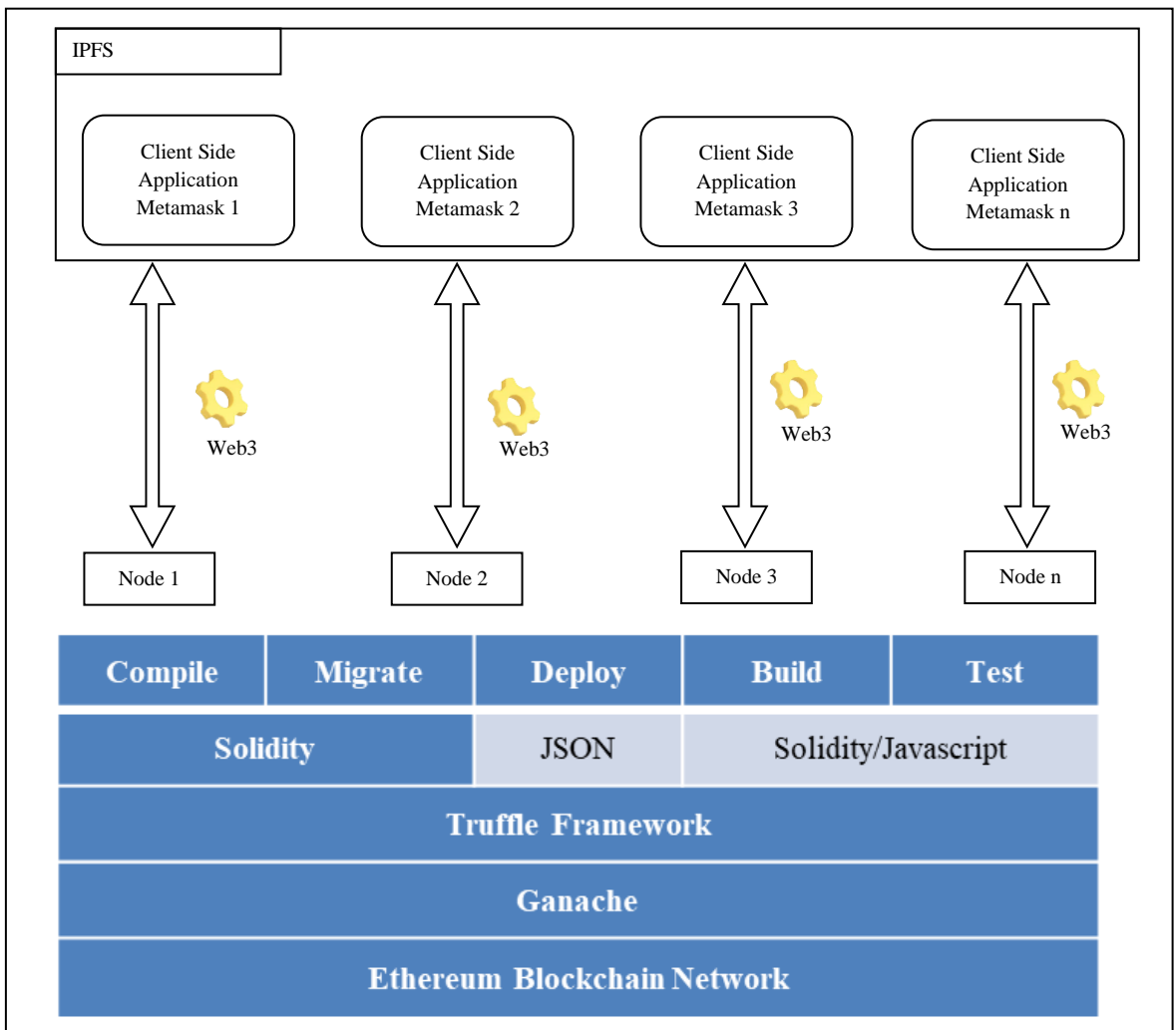

**Fig. 12 Ethereum Blockchain Network**

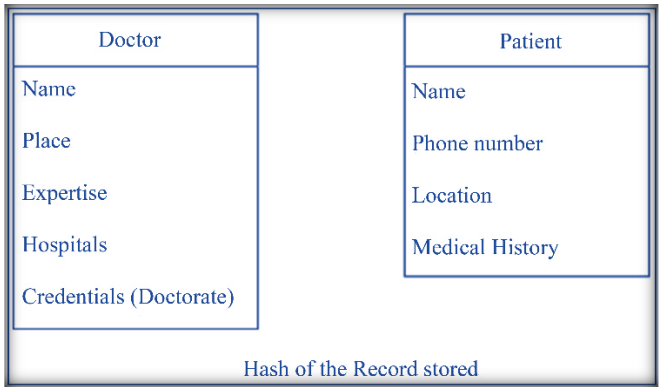

**Fig. 14 Models that get stored in Blockchain**

### 3.5. Blockchain and Machine Learning

Emergency clinic enrollment is made in light of the information of the medical clinic like emergency clinic name, mail id, telephone number, address, year of foundation, and secret word. The whole information is put away in a decentralized data set called Moralis. Approved enlisted emergency clinics register patients and specialists thinking about vital certifications. Blockchain, a decentralized and dispersed record, allows the enlisted emergency clinics to access the required patients' information. Patient certifications include the patient's name, telephone number, area, aadhar number, email id, clinical history, side effects, and age. Certifications for specialists incorporate the name, telephone number, area, email id, enlistment number, enrollment gathering, specialization, and experience. The whole information is put away in the Blockchain in a decentralized and got way. The specialists enrolled by each emergency clinic are apparent across the organization, and the patient information is available to the approved specialist. This demonstrates the decentralized property of Blockchain.

The information entered in the frontend part will be put away utilizing savvy contracts composed independently for specialists and patients. Metamask affirmation is obligatory for enlistment, and a specific number of ethers would be deducted for each exchange. Not long after showing specialists to all clinics then, any persistent who needs to counsel a specialist needs to counsel an emergency clinic administrator to such an extent that he/she will demand a specialist for an interview. When a specialist acknowledges the solicitation, he/she can investigate patients' clinical history and treat them likewise. After treatment, the specialist can send a solution to the patient who counseled him. Later patients can see the solution given. All the while, the patient can transfer his past clinical history report. Each specialist and patient goes about as a hub.

Here decentralization is accomplished to such an extent that each clinic can get to the enlisted specialists and counsel them right away. Suppose any difficult issue occurs with a patient. In that case, he can go to any neighboring emergency clinics and seek his therapy since the patient's clinical history is made effectively accessible through decentralization. In AI, separate models are made for significant illnesses like Heart, Diabetes, Kidney, Cellular breakdown in the lungs, and maternal well-being. Specialists will login to this page, and as indicated by the tests played out, the specific model will be gotten to. In light of the arrangement of side effects, prepared models will anticipate the seriousness of the sickness. These AI models with high precision will diminish the human reliance, exertion, and time required. Reports will be created once the expectation is finished. Specialists can create the reports for additional cycles.

## 4. Results

The UI of the clinical information executives framework will have five exercises, one for the information proprietor (the patient) and one more for the information requestors (specialist, clinic, path labs, and drug stores).

Every action will have enlistment and sign-in functionalities. During the hour of enlisting, every element will be given a remarkable id which will be referred to while getting to Blockchain. The patient will have a page where they can see their clinical records. Then again, different elements (specialists, emergency clinic, path lab, and drug stores) will approach see as well as transfer the assigned subtleties as another block to the Blockchain. To give a faultless encounter to the clients, a versatile application will be created utilizing a cross-stage instrument named Shudder. This will ensure that the client effort is not simply restricted to Android or iOS clients yet to both. The functioning framework is shown in Figure 15; it is the UI for our patients to enlist themselves if they are new to the site and refresh their new clinical well-being records to the data set through their ID and secret key. The medical clinic substance is where patients update their most recent experimental outcomes or medicine for any further remarks from the specialist allocated. As displayed in Figure 15, it is the pharmacy and lab segment where the particular offices' capabilities concur. The drug store gives the medication according to the solution, and the labs perform the tests referenced in the submitted structure.

### 4.1. Secure and Enhanced Protocol for the Storage of Private Health Data

Executing a Blockchain information structure gives changelessness of information and genuineness to the clients getting to the information. The information put away must be refreshed or embedded by the power that claims the confidential key while others could get to their information for the survey. This guarantees the genuineness that main approved individuals like clinics, guardians, and research facilities could refresh a patient's information. The blocks in the information tie are connected to past blocks utilizing the block's hash. Subsequently, any block adjustment should be refreshed in each block to gain an upgraded secure convention for the capacity of private well-being information.

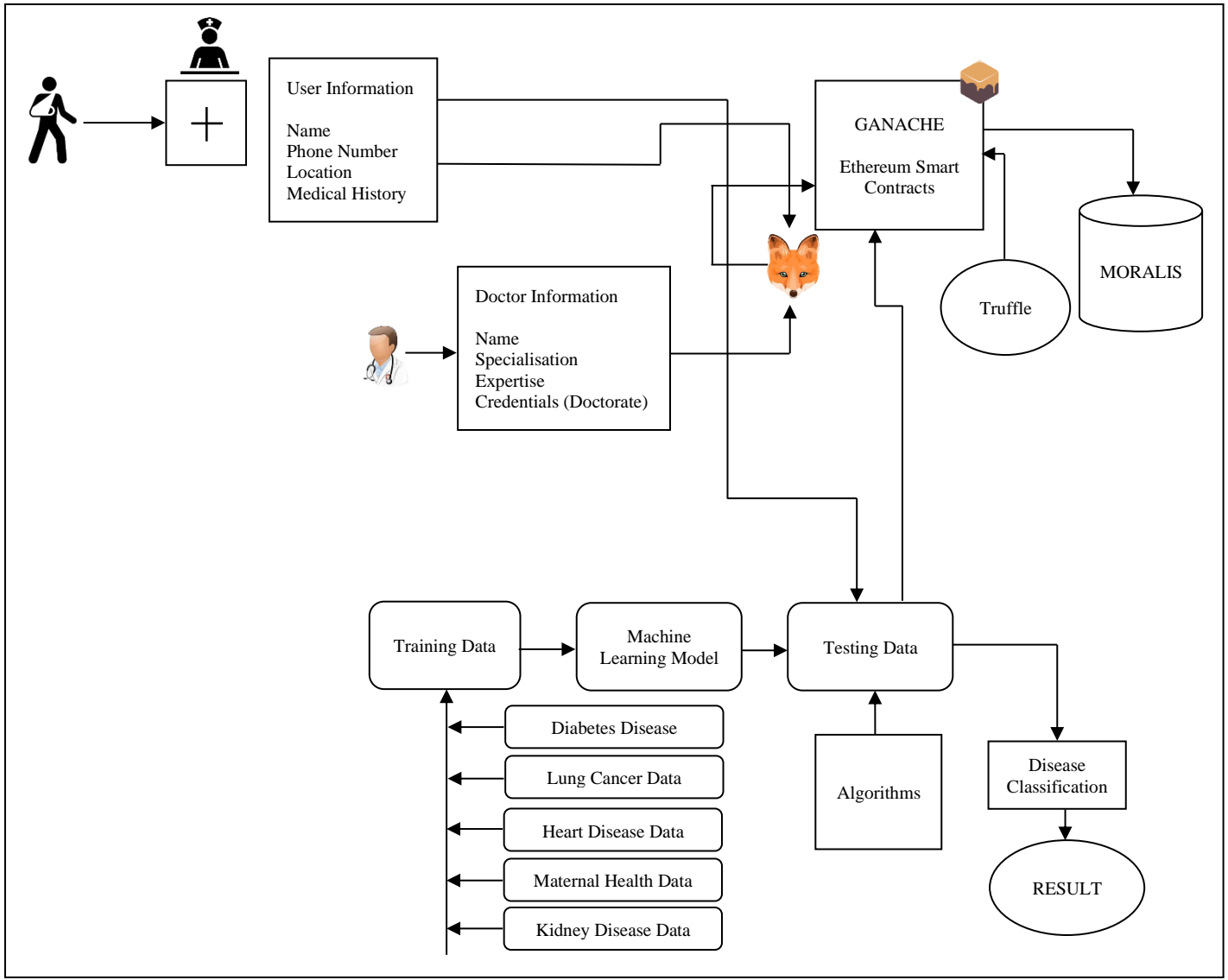

**Fig. 13 Workflow**

Each block holds a duplicate of the whole Blockchain. This makes our undertaking a safe construction to store the clinical information of every person.

### 4.2. An Economically Feasible Health Data Management System

Keeping a whole data set that incorporates all clinical records of every patient requires a lot of room which straightforwardly requires a tremendous measure of capacity, subsequently expanding the expense of keeping up with the Blockchain exponentially. To decrease the expense, the option to give the information requesters (the clinic and specialist) impermanent access was tried utilizing Mailman, as displayed in Figure 18. The progressions in the Blockchain can be pictured utilizing Hyperledger-texture voyager and securing the information inside various channels

The current arrangement of medical care is unified and not secure. At the same time, the proposed framework gives decentralization and gives suitable measures of safety to the pivotal clinical information utilizing agreement components. The decentralization helps entrance clinical information from any area of the planet, which would help in taking care of information misfortune issues during movement and absence of contribution during treatment.

- Different sicknesses per the arrangement of side effects are precisely anticipated by utilizing prepared AI models in the beginning phase and classified to assist scientists with monitoring new variations.
- Guarantee that the prepared model channels the information, disposes of the multitude of individual subtleties and data of the patient, and safeguards it from information holes and breaks.

- Power the framework with Blockchain to get the exchanges among patients and the approved specialist and give patients the adaptability to pick of capacity and give monetarily feasible well-being information the executive's framework, we have kept an off-stockpiling data set where we keep the genuine information of every person. However, the real Blockchain holds the connection to the distributed storage where the clinical resources are put away.

After the execution of the arrangement proposed in the paper, we had.

- Which specialist ought to see his well-being record, and who should not?
- Machine Learning Predictors for Lung Cancer, Diabetes, Heart Disease, Kidney Disease, and Maternal Health. Blockchain smart contracts are compiled and deployed in local Blockchain ganache.

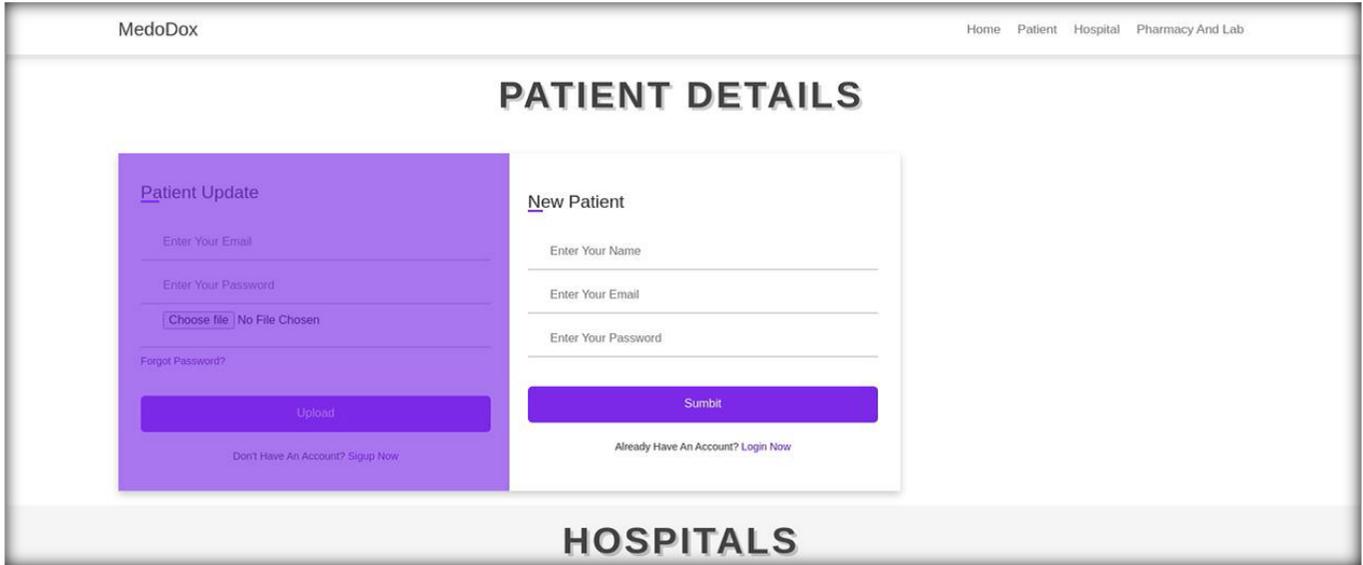

**Fig. 15 UI for patient registration**

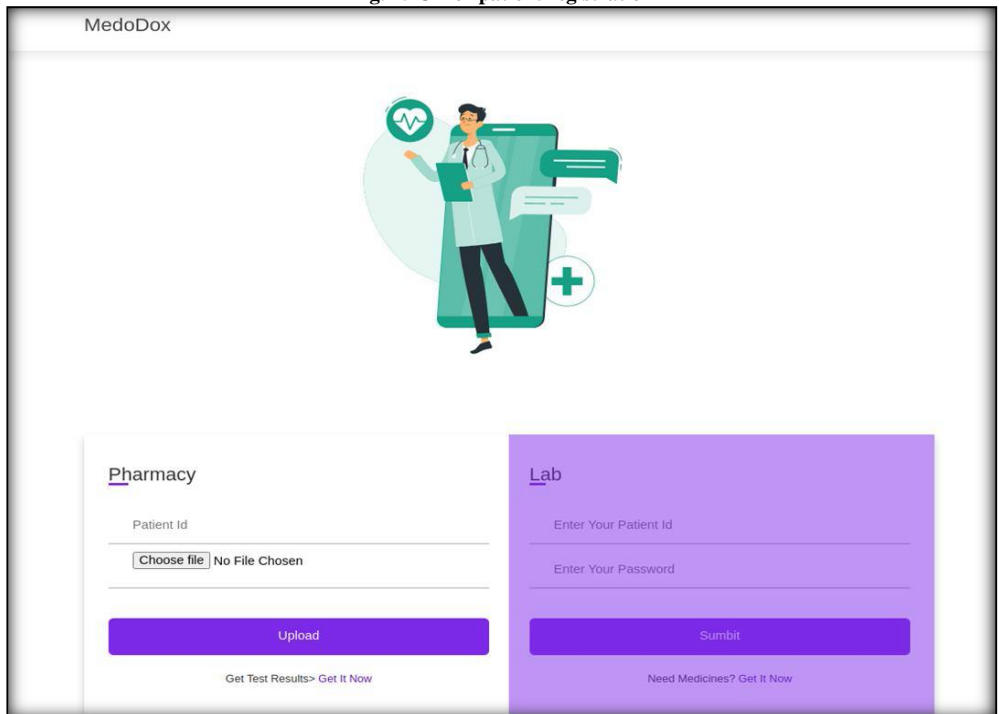

**Fig. 16 UI for Updation in Patient Blockchain by Pharmacy and Lab**

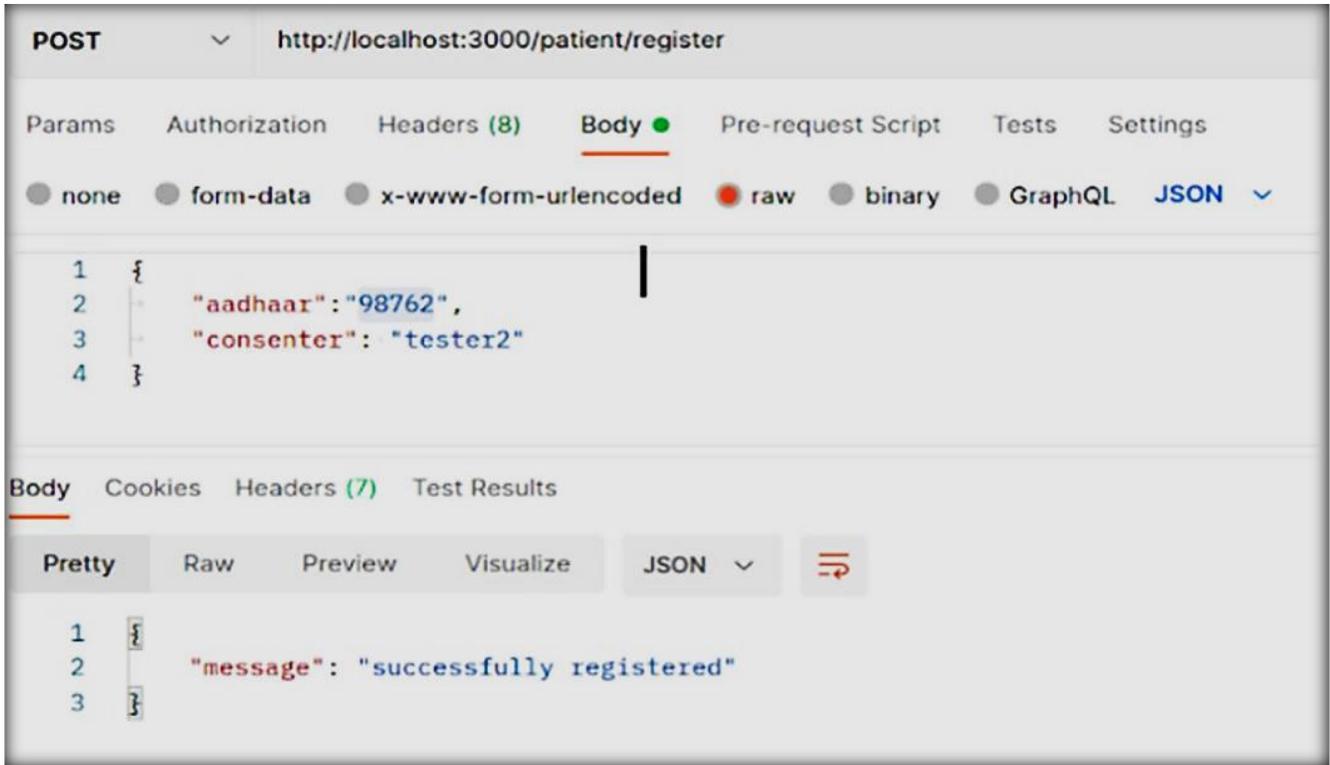

**Fig. 17 API response for a patient entity**

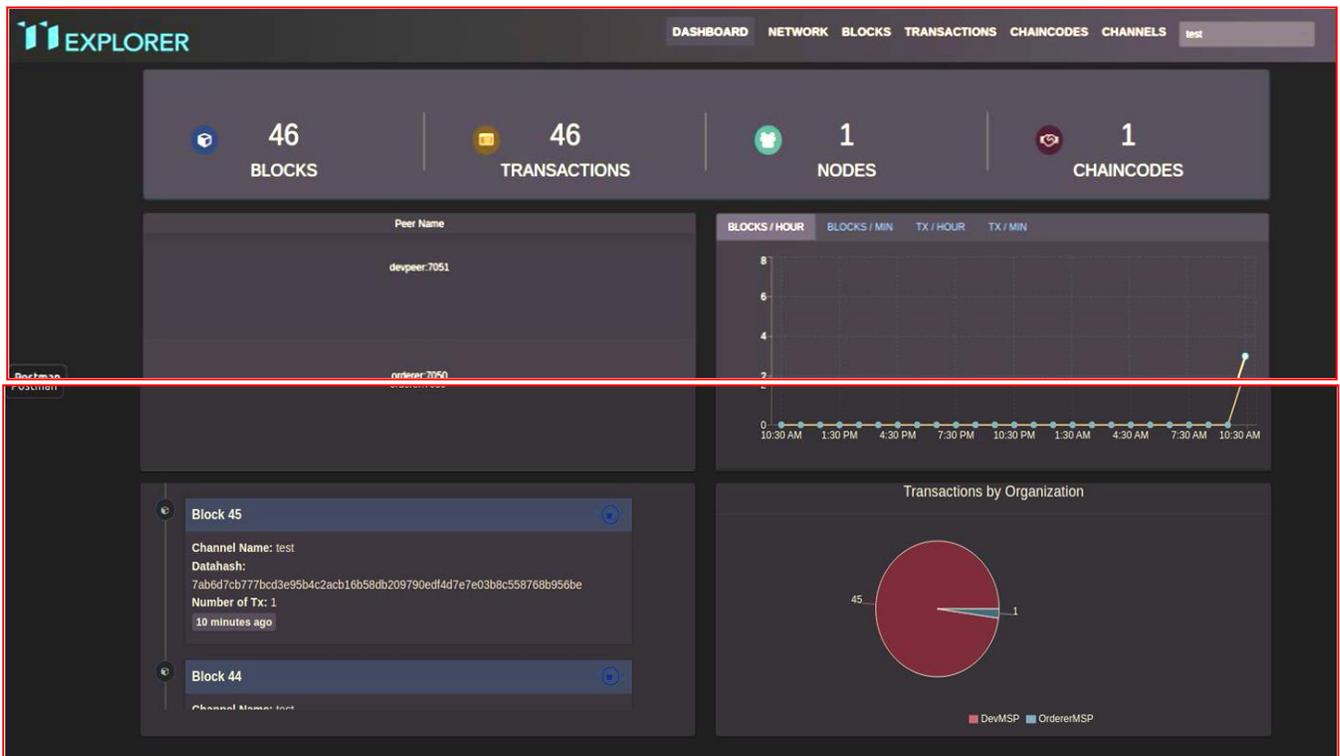

**Fig. 18 API response for a patient entity**

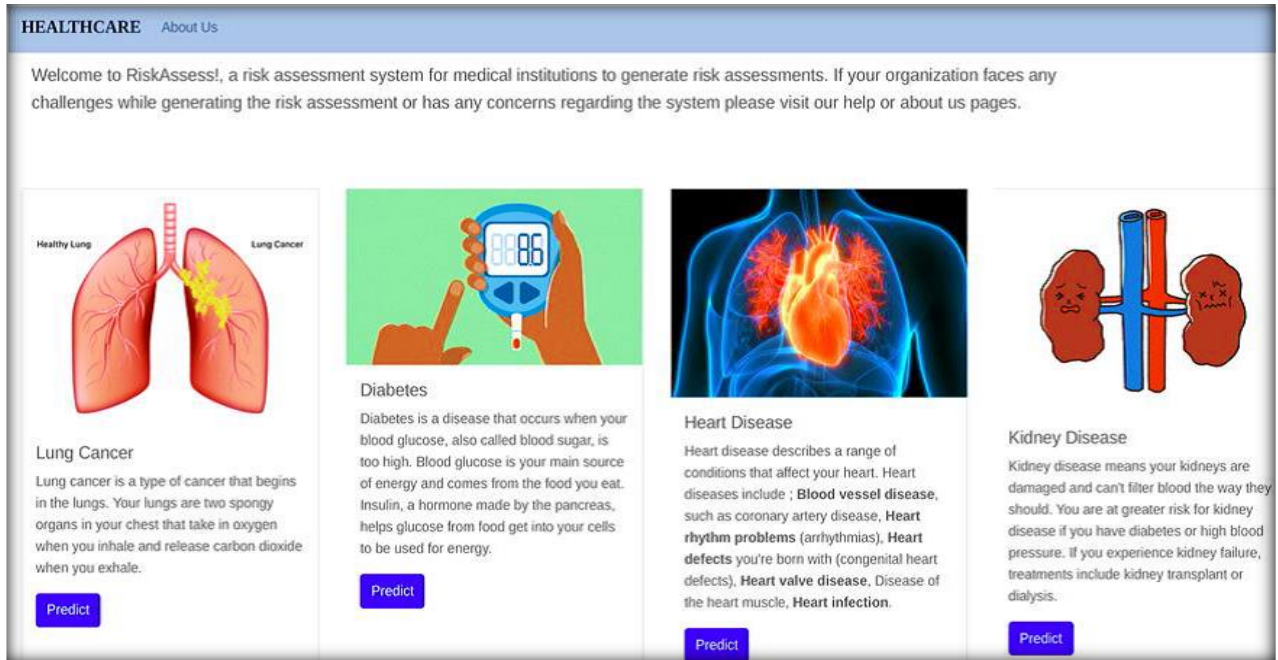

**Fig. 19 Home Screen**

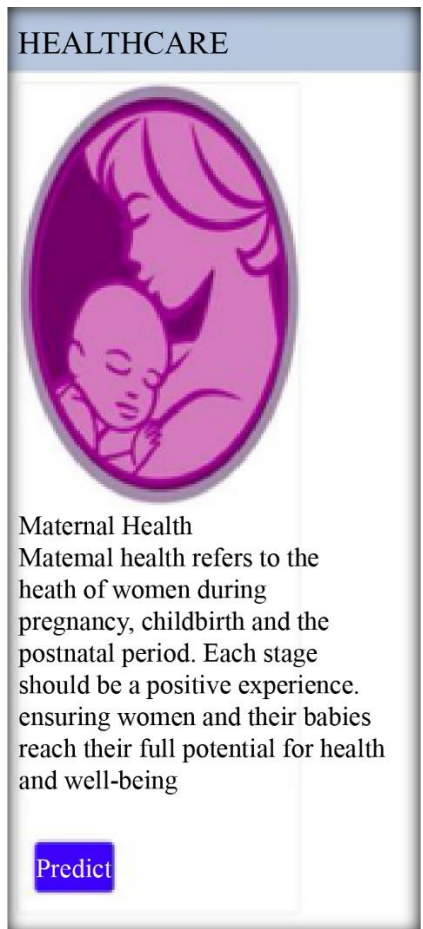

**Fig. 20 Home Screen**

| | |
|---|---|
| gender | 0/1 |
| Age in Years | eg. 34 |
| smoking | 1-no.2-yes |
| yellow_fingers | 1/2 |
| Anxiety | 1/2 |
| peer pressure | 1/2 |
| chronic disease | 1/2 |
| fatigue | 1/2 |
| allergy | 1/2 |
| wheezing | 1/2 |
| alcohol consuming | 1/2 |
| coughing | 1/2 |
| shortness of breath | 1/2 |
| swallowing difficulty | 1/2 |
| chest pain | 1/2 |

Predict

**Fig. 21 Lung Cancer Predictor**

**Fig. 22 Kidney Disease Predictor**

**Fig. 24 Maternal Health predictor**

**Fig. 23  Diabetes Predictor**

**Fig. 25 Heart Disease Predictor**

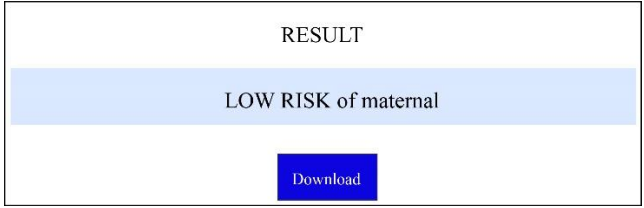

**Fig. 26 Predictor Result**

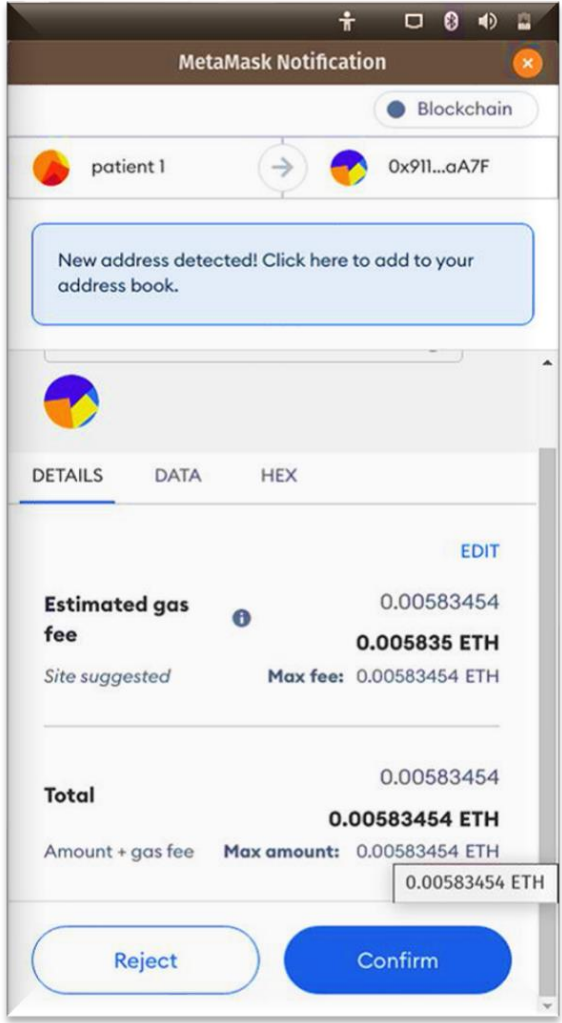

**Fig. 27 Metamask Confirmation for a transaction**

## 5. Conclusion

As the world pushes ahead, each part of life is estimated concerning significant investment utilization. The clinical framework should be redesigned to stay aware of business as usual. One type of medical services overhaul is to work with effective access and recovery of information resources between patients, specialists, and parental figures while parallelly keeping a protected convention for the openness of private information resources. This paper contributes to giving a helpful approach to facilitating the recovery and capacity of a singular's well-being record utilizing a Blockchain information structure which makes information put away in each block to be permanent. In the Blockchain, the blocks are connected to the past block with hash esteem, and each block has a duplicate of the whole Blockchain. In this manner, any change or update on any block should be pondered each block, making it computationally difficult to penetrate and bring about Blockchain to be an extremely gotten information construction to store clinical records.

Limiting who can refresh or add information to the Blockchain likewise helps in giving security or obstruction from any kind of pantomime. This is worked by utilizing public and confidential keys, where the confidential keys limit just approved clients to refreshing the records. Emergency clinics and specialists are permitted to refresh the experimental outcomes, suggestions, or treatment examination in the clinical information resource.

At last, this paper means to assume a significant part in the medical services and the executives of information resources for patients to have the option to refresh their new reports and get the required report examination and remedies from the clinical specialists while likewise guaranteeing that the respectability of their information is kept up with. Blockchain innovation has been created with time, and banking and money-related regions are presently using Blockchain, keeping up with as an essential concern the unmatched advantages it offers. With the basic development in prosperity, data get through hacking, and utilization of Blockchain for security becomes critical and basic.

AI, on the other hand, has transformed into a necessary development vital due to the various regions it has, at this point, benefited. The wide extent of areas consolidates cultivating, vehicle industry, and medical services industries. The goal is to confirm one estimation that can yield the best exactness. Involving ML estimations in medical care structures would clear a technique for expecting and thwarting diseases in their slow stages and help with saving lives. It will not be inappropriate to say that the Blockchain-based Medical services model is the future in the medical services area and can have an impact on how medical services records are overseen and gotten. With the improvement of 5-G associations and speedier than at some other times data move workplaces, it will empower the progress of AI, Blockchain, and different data-based techniques in various regions, including Medical Services. As this new creative natural framework emerges, Blockchain ensures basic updates regulating patient prosperity records. Tireless undertakings are being made to fabricate the accuracy of wearable prosperity GPS guides. Suppose these data could give more definite and reliable results. In that case, there will be more splendid conceivable outcomes planning these contraptions with the prosperity records to give more information and securely share a piece of these clinical data with endorsed experts without truly visiting.

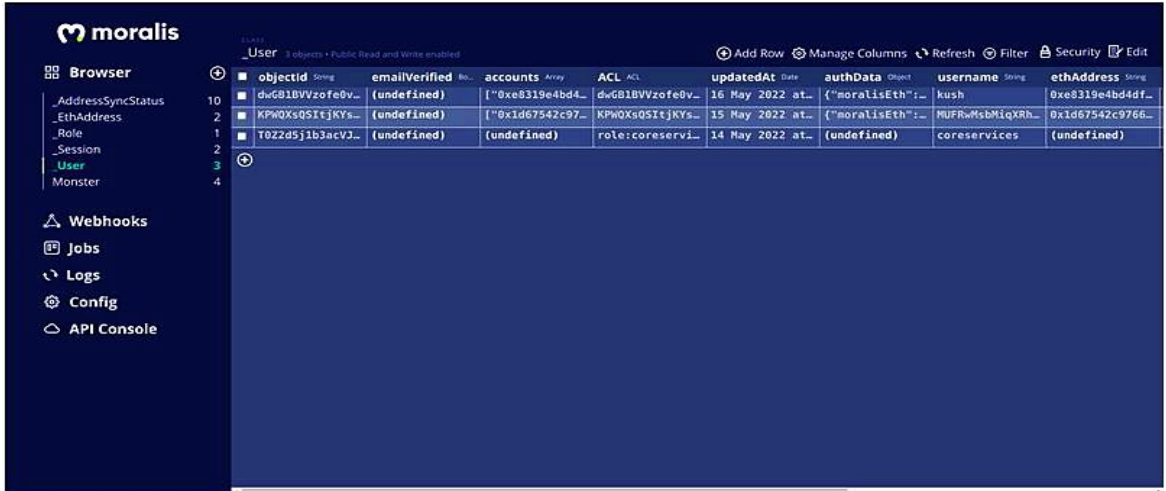

**Fig. 28 Storing Data in Offchain Moralis**

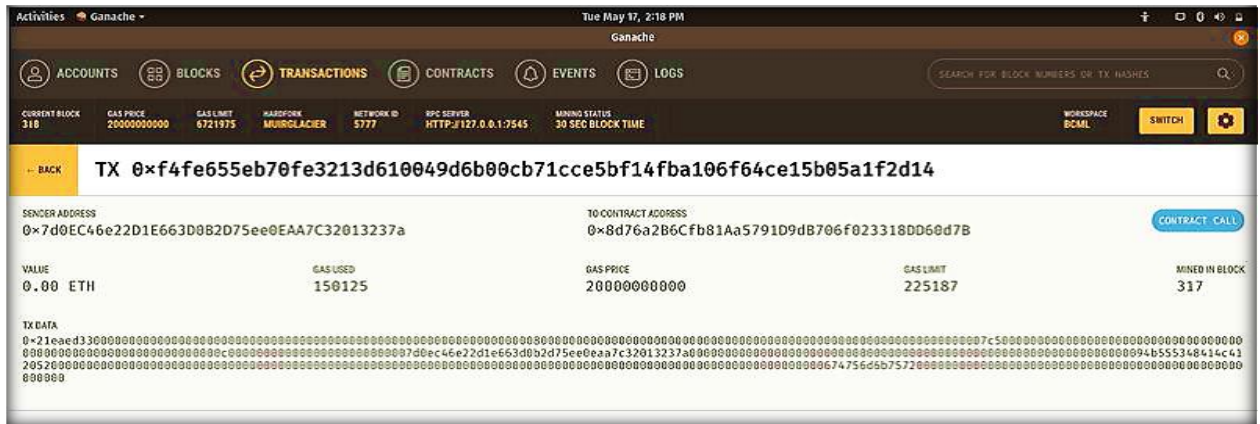

**Fig. 29 Transaction in Ganache**

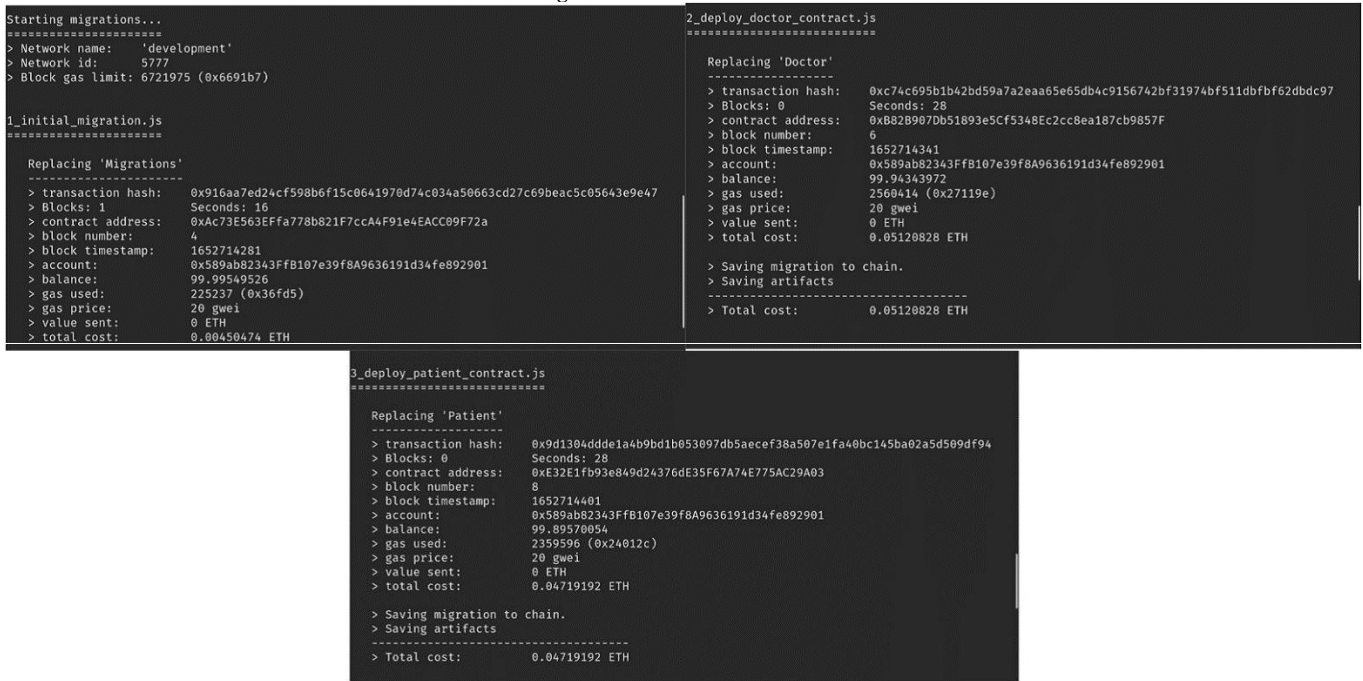

**Fig. 30 Deploying contracts in Blockchain**

## Acknowledgments

I would like to express my heartfelt gratitude to our supervisor, Dr. Poornima A.S., for her unwavering support and guidance throughout the research process. Her expertise, insights, and encouragement were invaluable in helping us to complete this work. Finally, I would like to thank my family and friends for their love and support throughout the research process. Without their encouragement and support, I would not have been able to complete this research.